\documentclass[%
  twocolumn,%
  groupedaddress,%
  showpacs,%
  showkeys,%
  prb,%
  letterpaper,%
  amsfonts,%
  floatfix,%
]{revtex4}

\usepackage{amsmath,amssymb,bm}
\usepackage[pdftex]{graphicx}
\usepackage{subfigure}
\usepackage{hyperref}
\hypersetup{%
  breaklinks = {true},
  citecolor = {blue},
  colorlinks = {true},
  linkcolor = {red},
  pdfauthor = {\textcopyright\ Vitor M. Pereira},
  pdfcreator = {\LaTeX\ and \flqq hyperref\frqq},
  pdffitwindow = {true},
  pdfmenubar = {true},
  pdfpagelayout = {SinglePage},
  pdfstartview = {Fit},
  pdftoolbar = {true},
  plainpages = {false},
}

\newcommand{\bk}{\ensuremath{\bm{k}}}
\newcommand{\bd}{\ensuremath{\bm{\delta}}}

\newcommand{\bR}{\ensuremath{\bm{R}}}

\newcommand{\ce}{\ensuremath{\varepsilon}}

\newcommand{\etal}{\emph{et al.}}
\newcommand{\Fref}[1]{Fig.~\ref{#1}}

\newcommand{\zdir}{\ensuremath{\mathcal{Z}}}
\newcommand{\adir}{\ensuremath{\mathcal{A}}}

\begin{document}


\title{A tight-binding approach to uniaxial strain in graphene}

\pacs{81.05.Uw,62.20.-x,73.90.+f}

%

\keywords{graphene, strain, elasticity, electronic structure, tight-binding}

\author{Vitor M. Pereira}
\affiliation{Department of Physics, Boston University, 590
Commonwealth Avenue, Boston, MA 02215, USA}

\author{N. M. R. Peres}
\affiliation{Centro de F\'{\i}sica  e  Departamento de
F\'{\i}sica, Universidade do Minho, P-4710-057, Braga, Portugal}

\author{A.~H. Castro Neto}
\affiliation{Department of Physics, Boston University, 590
Commonwealth Avenue, Boston, MA 02215, USA}

\date{\today}


\begin{abstract}
We analyze the effect of tensional strain in the electronic
structure of graphene. In the absence of electron-electron interactions, 
within linear elasticity theory, and a tight-binding approach, 
we observe that strain can generate a bulk spectral gap.
However this gap is critical, requiring threshold deformations in excess of
$20$\%, and only along preferred directions with respect to the underlying 
lattice. The gapless Dirac spectrum is robust for small and moderate
deformations, and the gap appears as a consequence of the merging of the two
inequivalent Dirac points, only under considerable deformations of the lattice.
We discuss how strain-induced anisotropy and local deformations can be
used as a means to affect transport characteristics and pinch off current flow
in graphene devices.
\end{abstract}

\maketitle


%
%
%
It is now well established that $sp^2$ bonded carbon systems feature
record-breaking mechanical strength and stiffness. Investigations in
the context of carbon nanotubes reveal intrinsic strengths 
\cite{Yu:2000} that make these systems the strongest in nature, 
Recently, graphene --- the mother of all $sp^2$ carbon structures --- 
has been confirmed as the strongest material ever to be measured
\cite{Lee:2008}, being able to sustain reversible elastic deformations in excess
of 20\% \cite{Liu:2007}.

These mechanical measurements arise at a time where graphene draws considerable
attention on account of its unusual and rich electronic properties. 
Besides the great crystalline quality, high mobility and resilience to high
current densities \cite{Geim:2007}, they include a strong field effect
\cite{Novoselov:2004}, absence of backscattering \cite{Ando:1998b} and a
minimum metallic conductivity \cite{Novoselov:2005}. While many such 
properties might prove instrumental if graphene is to be used in future
technological applications in the ever pressing demand for miniaturization in
electronics, the latter is actually a strong deterrent: it hinders the
pinching off of the charge flow and the creation of quantum point contacts. 
In addition, graphene has a gapless spectrum
with linearly dispersing, Dirac-like, excitations
\cite{Wallace:1947,CastroNeto:2007}. Although a gap can be induced by means of
quantum confinement in the form of nanoribbons \cite{Han:2007} and quantum dots
\cite{Ponomarenko:2008}, these ``paper-cutting'' techniques are prone to edge
roughness, which has detrimental effects on the electronic properties.
Hence, a route to induce a robust, \emph{clean}, bulk spectral gap in
graphene is still much in wanting. 
 
In this paper we inquire whether the seemingly independent aspects
of mechanical response and electronic properties can be brought together with
profit in the context of a tunable electronic structure. Motivated by 
recent experiments showing that reversible and controlled strain can be
produced in graphene with measurable effects \cite{Ni:2008b,Ni:2008a}, we
theoretically explore the effect of strain in the electronic structure of
graphene within a tight-binding approach. 
Our calculations show that, in the absence of electron-electron interactions, a
gap can be opened in a pure tight binding model of 
graphene for deformations beyond 20\%. This gap opening is not a consequence
of a broken sublattice symmetry but due to level crossing. 
The magnitude of this effect depends on the direction of applied tension, so
that strain along a zig-zag direction is most effective in overcoming the gap
threshold, whereas deformations along an armchair direction do not induce a
gap. Unfortunately, such large threshold deformations render strain an
ineffective means to achieve a bulk gapped spectrum in graphene. We discuss
alternate means to impact transport and electronic structure using local strain
profiles.

%
%
\begin{figure}[b]
  \centering
  \subfigure[][]{
    \includegraphics[width=0.8\columnwidth]{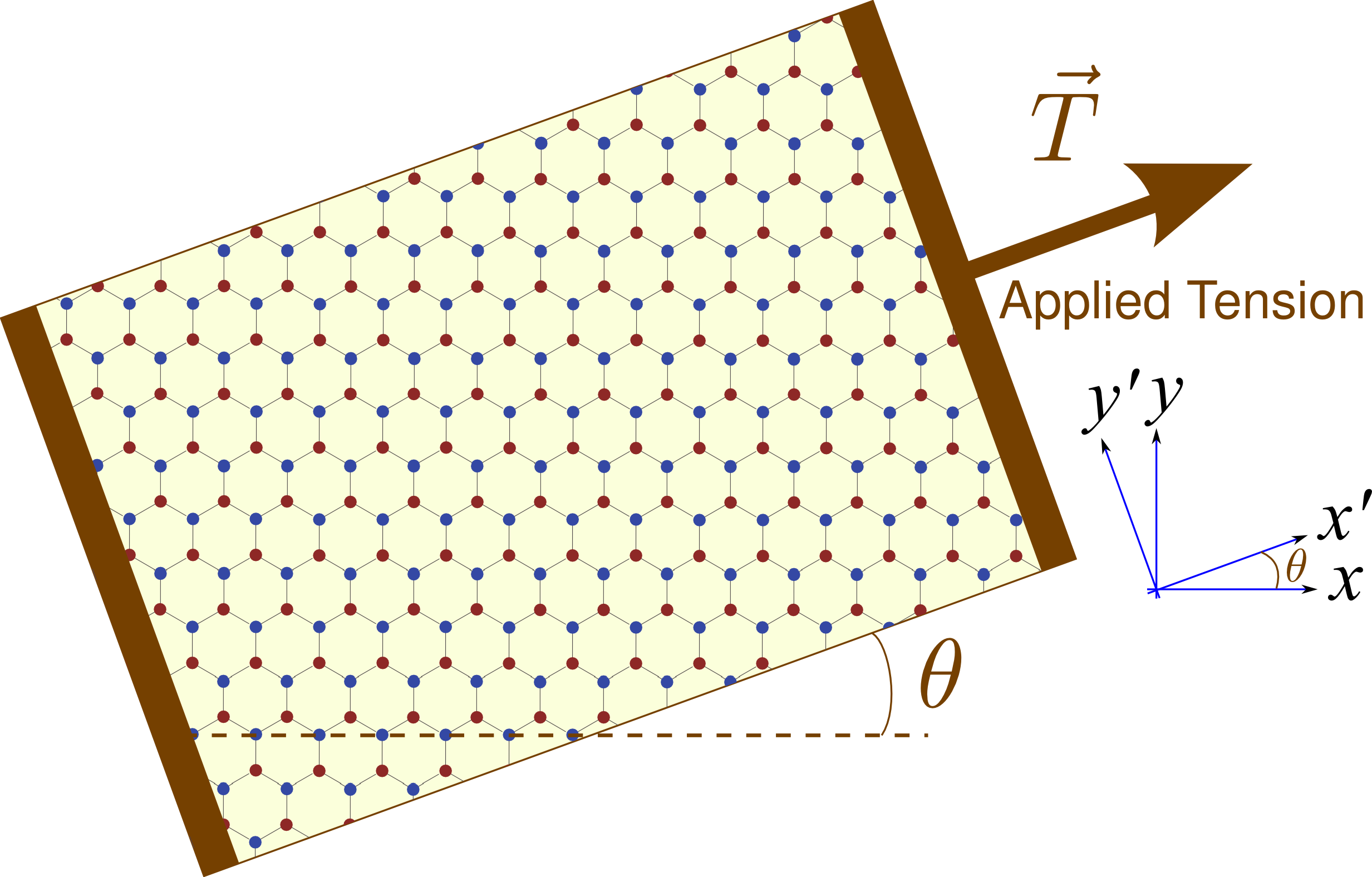}  
    \label{fig:Geometry}
  }
  \caption{
    (Color online)
    \subref{fig:Geometry} Tension geometry considered in the text. The zig-zag
    direction of the honeycomb lattice is always parallel to the axis $Ox$.
  }
  \label{fig:Scheme}
\end{figure}
%

%
%
\begin{figure}[t]
  \centering
  \subfigure[][]{
    \includegraphics[width=0.4\columnwidth]{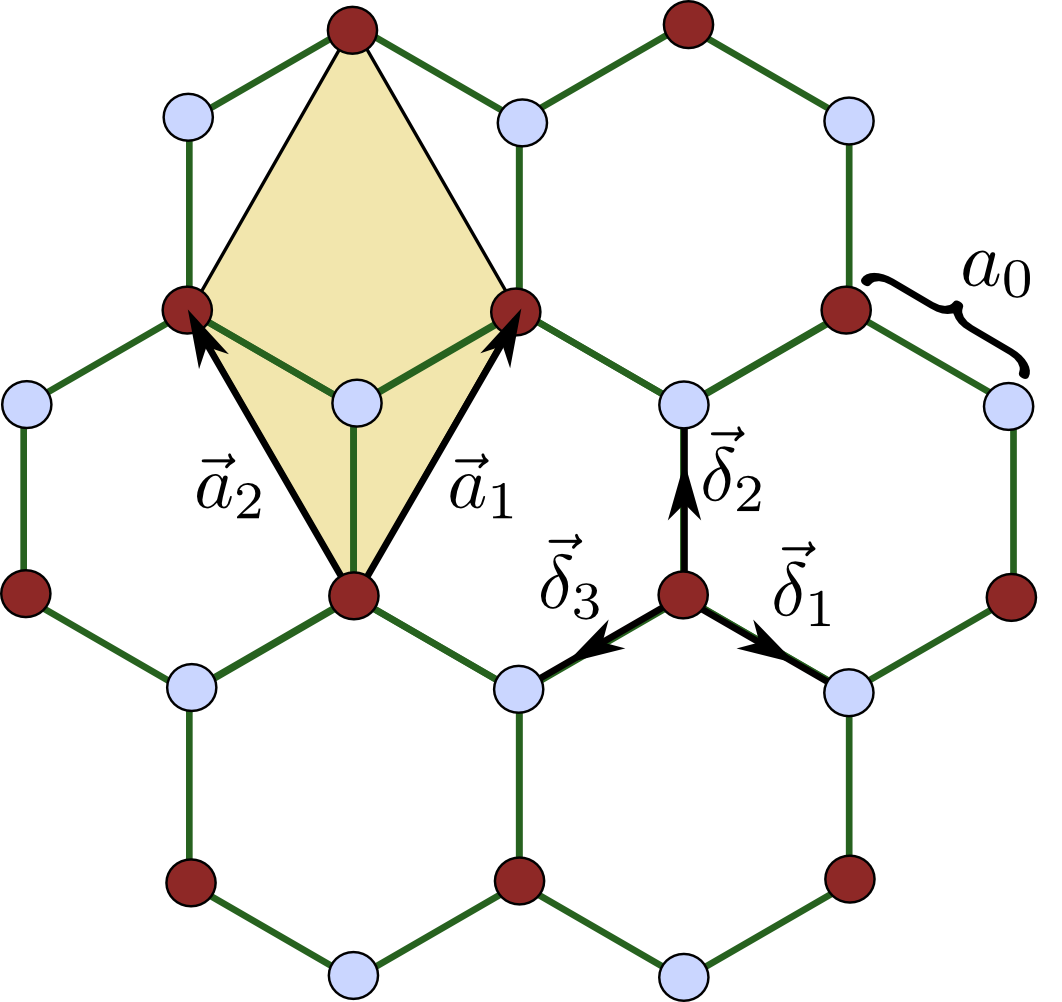}  
    \label{fig:Lattice}
  }
  \quad
  \subfigure[][]{
    \includegraphics[width=0.4\columnwidth]{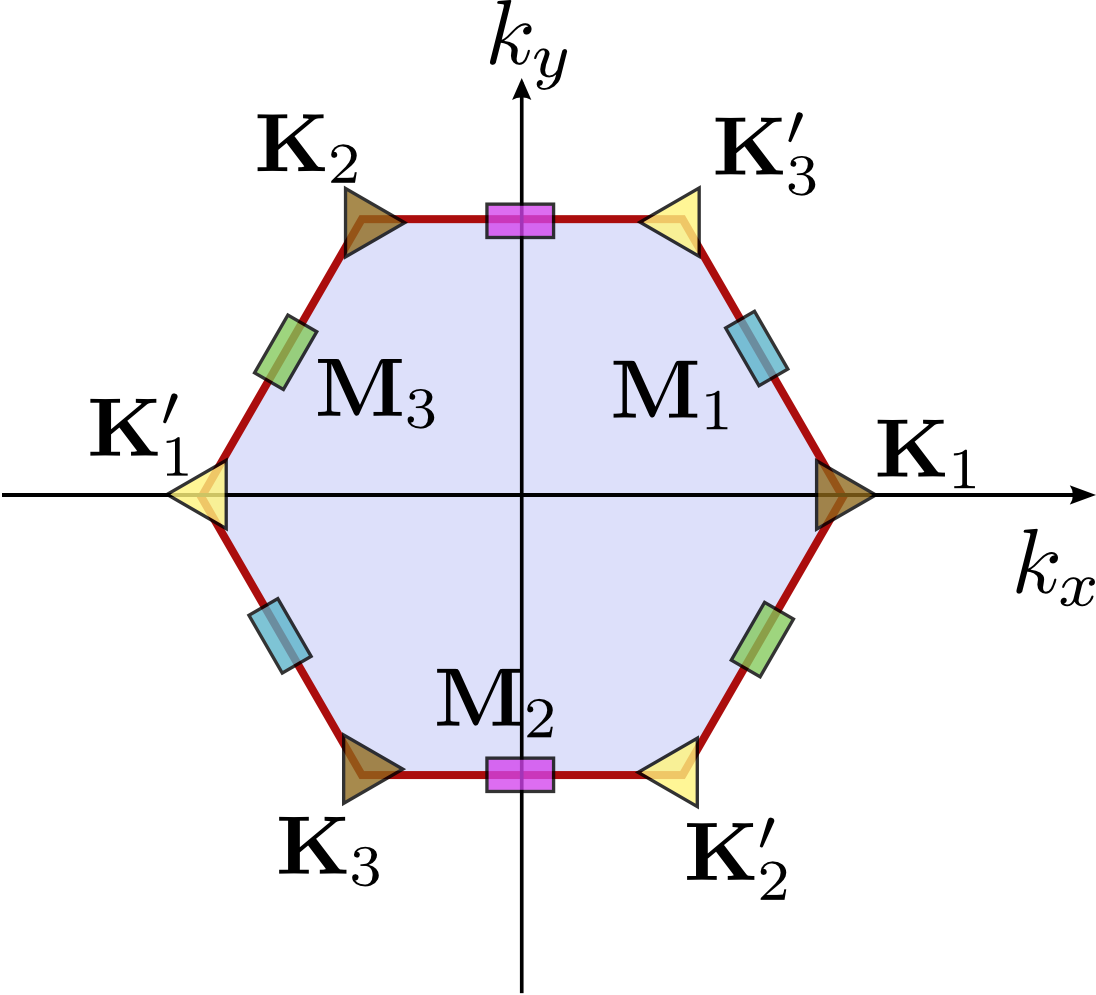}  
    \label{fig:Reciprocal}
  }
  \caption{
    (Color online)
    \subref{fig:Lattice} Honeycomb lattice geometry. The vectors
    $\bd_1=a(\tfrac{\sqrt{3}}{2},-\tfrac{1}{2})$, $\bd_2=a(0,1)$
    $\bd_3=a(-\tfrac{\sqrt{3}}{2},-\tfrac{1}{2})$ connect A-sites (red/dark) to
    their B-site (blue/light) neighbors.
    \subref{fig:Reciprocal} The first Brillouin zone of undeformed graphene,
    with its points of high symmetry.
  }
  \label{fig:Lattices}
\end{figure}
%

%
%
\section{Model}
We consider that electron dynamics of electrons hopping in the honeycomb      
lattice is governed by the nearest neighbor tight-binding Hamiltonian         
\begin{equation}                                                              
  H = \sum_{\bR,\bd} t(\bR,\bd) a^\dagger(\bR)b(\bR+\bd) + \text{H. c.}       
  \,.                                                                         
  \label{eq:Hamiltonian}                                                      
\end{equation}                                                                
Here $\bR$ denotes a position on the Bravais lattice, and $\bd$ connects the  
site $\bR$ to its neighbors; $a(\bR)$ and $b(\bR)$ are                        
the field operators in sublattices A and B. The first thing to emphasize is   
that, under general stress conditions, the hopping $t(\bR,\bd)$ will be     
generally different among different neighbors. We are interested in the
elastic                                                                         
response, for which deformations are affine. This means that even though the  
hoppings from a given atom to its neighbors can be all different, they will be
the same for every atom. Therefore, as depicted in \Fref{fig:Lattice}, we need
only to consider three distinct hoppings: $t_1=t(\bd_1)$, $t_2=t(\bd_2)$,     
and $t_3=t(\bd_3)$. The relaxed equilibrium value for $t_i=t(\bd_i)$ is         
$t_0\approx 2.7\;\text{eV}$ \cite{CastroNeto:2007}. Our goal is to investigate
the changes that strain induces in these hoppings, and what impact they have in
the resulting electronic structure.                                           

Throughout this paper we shall use the C--C equilibrium distance,
$a_0=1.42\text{
\AA}$, as unit of length, and will frequently use $t_0$ as unit of energy.

%
%
\section{Analysis of Strain}
We are interested in uniform planar tension situations, like the one
illustrated in \Fref{fig:Geometry}: the graphene sheet is uniformly stretched
(or compressed) along a prescribed direction. The fixed Cartesian system is
chosen in a way that $Ox$ always coincides with the zig-zag direction of the
lattice. In these coordinates the tension, $\bm{T}$, reads $\bm{T} = T
\cos(\theta)\,\bm{e}_x + T \sin(\theta)\,\bm{e}_y$.

As for any solid, the generalized Hooke's law relating stress, $\tau_{ij}$ and
strain $\ce_{ij}$ has the form
\begin{equation}
  \tau_{ij} = C_{ijkl} \, \ce_{kl}
  \,,\quad
  \ce_{ij} = S_{ijkl} \, \tau_{kl}  
\end{equation}
where $C_{ijkl}\,(S_{ijkl})$ are the components of the stiffness (compliance)
tensor. Since we address only states of planar stress, we resort to the
2-dimensional reduction of the stress and strain tensors. In general the
components $C_{ijkl}$ depend on the particular choice of the Cartesian axes.
Incidentally, for an hexagonal system under planar stress in the basal plane,
the elastic components are independent of the coordinate system. This means that
graphene is elastically isotropic \cite{Landau-Elasticity}.

The analysis of strain is straightforward in the principal system $Ox'y'$ where 
we simply have $\bm{T} = T \bm{e}_{x'}$:
\begin{equation}
  \ce_{ij}'=S_{ijkl}\tau_{kl}'=T\, S_{ijkl}\delta_{kx}\delta_{lx}=T\,S_{ijxx}
\end{equation}
Given that only five compliances are independent in graphite
(viz., $S_{xxyy},\, S_{xxyy},\, S_{xxzz},\, S_{zzzz},\, S_{yzyz}$)
\cite{Blakslee:1970}, it follows that the only non-zero deformations are
\begin{equation}
  \ce_{xx}'=T\, S_{xxxx}\,,\:\ce_{yy}'=T\, S_{xxyy}
  \,,
\end{equation}
which represent the longitudinal deformation and Poisson's transverse
contraction. If we designate the tensile strain by $\ce=T\, S_{xxxx}$, the
strain tensor can be written in terms of Poisson's
ratio, $\sigma=-S_{xxxy}/S_{xxxx}$:
\begin{equation}
  \bm{\ce}' = \ce \begin{pmatrix}
                    1 & 0\\0 & -\sigma
                  \end{pmatrix}
  \,.
  \label{eq:PrincipalStrainTensor}
\end{equation}
This form shows that graphene responds as an isotropic elastic medium.
For Poisson's ratio we use the value known for graphite: $\sigma=0.165$
\cite{Blakslee:1970}. 
It should be mentioned that when stress is induced in graphene by mechanically
acting on the substrate (i.e. when graphene is adhering to the top of a
substrate and the latter is put under tension, as is done
in Ref.~\onlinecite{Ni:2008b}), the relevant parameter is in fact
the tensile strain, $\ce$, rather than the tension $T$ \cite{Endnote:2}.
For this reason, we treat $\ce$ as the tunable parameter.
Since the lattice is oriented with respect to the axes $Oxy$,
the stress tensor needs to be rotated to extract information about bond
deformations. The strain tensor in the lattice coordinate system reads
\begin{equation}
  \bm{\ce} = \ce 
  \begin{pmatrix}
     \cos^2\theta -\sigma\sin^2\theta & 
     (1+\sigma)\cos\theta\sin\theta \\
     (1+\sigma)\cos\theta\sin\theta &
     \sin^2\theta -\sigma\cos^2\theta
  \end{pmatrix}
  \,.
  \label{eq:StrainTensor}
\end{equation}
%

%
%
\begin{figure}[tb]
  \centering
  \subfigure[][]{%
    \includegraphics[width=0.62\columnwidth]{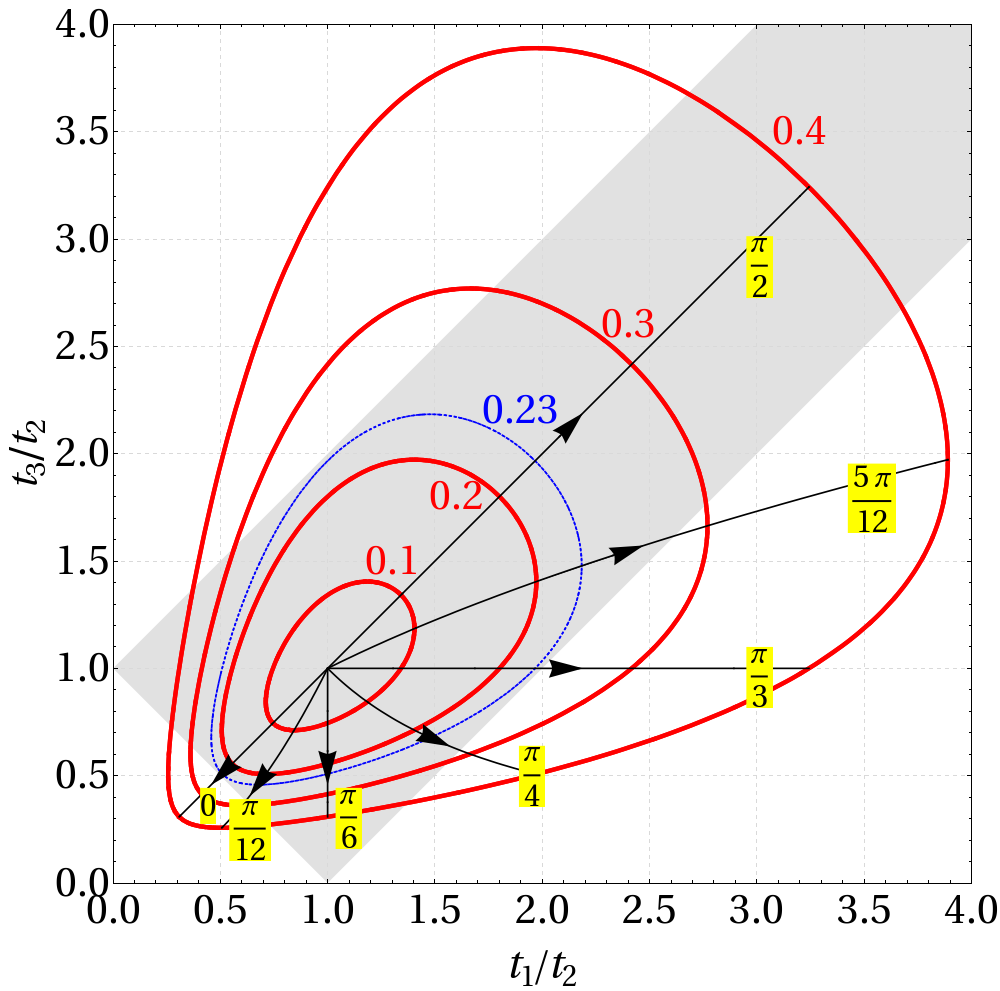}%
    \label{fig:GapThreshold}%
  }%
  \subfigure[][]{%
   
\includegraphics[width=0.38\columnwidth]{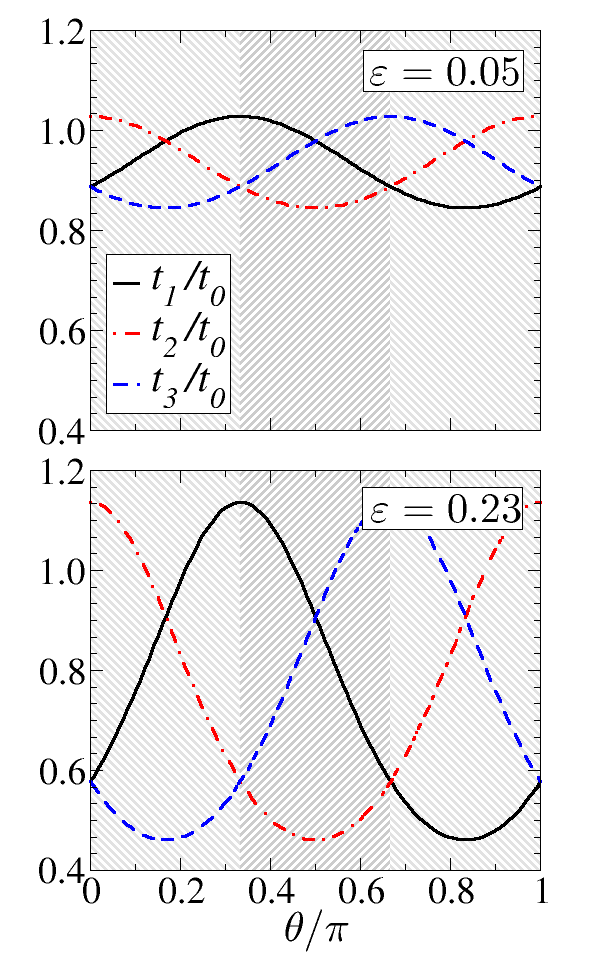}
    \label{fig:Hoppings}%
  }
  \caption{
    (Color online)
    \subref{fig:GapThreshold} Plot of $t_1/t_2$ vs $t_3/t_2$ as a function of 
    strain, $\ce$, and $\theta$. Closed lines are iso-strain curves, and
    arrowed lines correspond to the trajectory of the point
    $(t_1/t_2,t_3/t_2)$ as $\ce$ increases, calculated at constant
    angle. 
    The graph is symmetric under reflection on both axes.
    In the shaded area the spectrum is gapless. The blue
    iso-strain line ($\ce \approx 0.23$) corresponds to the gap threshold.
    In panel \subref{fig:Hoppings} we show the angular dependence of
    $t_{1,2,3}$ for $\ce=0.05$ and $\ce=0.23$.
  }
  \label{fig:StrainPlots}
\end{figure}
%

%
%
\section{Bond Deformations}
If $\bm{v}^0$ represents a general vector in the undeformed graphene
plane, its deformed counterpart is given, to leading order, by the
transformation
\begin{equation}
  \bm{v} = (\bm{1}+\bm{\ce})\cdot\bm{v}^0
  \,.
  \label{eq:Gen-Vec-Deformation}
\end{equation}
Especially important are the deformations of the nearest-neighbor bond
distances. Knowing $\ce_{ij}$ one readily obtains the deformed bond vectors
using \eqref{eq:Gen-Vec-Deformation}. The new bond lengths are then given by
\begin{subequations}\label{eq:DeformedBonds}
\begin{align}
  |\bd_1| &=
    1+\tfrac{3}{4}\ce_{11}-\tfrac{\sqrt{3}}{2}\ce_{12}+\tfrac{1}{4}\ce_{22}\\
  |\bd_2| &= 1+\ce_{22}\\
  |\bd_3| &=
    1+\tfrac{3}{4}\ce_{11}+\tfrac{\sqrt{3}}{2}\ce_{12}+\tfrac{1}{4}\ce_{22} 
  \,
\end{align}
\end{subequations}
Of particular interest are the cases $\theta=0$ and 
$\theta=\pi/2$ since they correspond to tension along
the zig-zag (\zdir) and armchair (\adir) directions:
\begin{subequations}\label{eq:DeformedBonds-ZA}
\begin{align}
  \zdir:\quad&
  |\bd_1| =|\bd_3| = 1 + \tfrac{3}{4}\ce -\tfrac{1}{4}\ce\sigma
  \,,\;
  |\bd_2| =1 - \ce\sigma
  \\
  \adir:\quad&
  |\bd_1| =|\bd_3| = 1 + \tfrac{1}{4}\ce -\tfrac{3}{4}\ce\sigma
  \,,\;
  |\bd_2| =1 + \ce
\end{align}
\end{subequations}

The modification of these distances distorts the reciprocal lattice as well,
and the positions of the high-symmetry points shown in \Fref{fig:Reciprocal} are
shifted. The primitive vectors of the reciprocal space are denoted by
$\bm{b}_{1,2}$, and, in leading order, change according to
\begin{subequations}\label{eq:Reciprocal-Vectors}
\begin{align}
  \bm{b}_1 \approx \frac{2\pi}{\sqrt{3}} 
    \left( 1 - \ce_{11} - \frac{\ce_{12}}{\sqrt{3}};\,
    \frac{1}{\sqrt{3}} - \ce_{12} - \frac{\ce_{22}}{\sqrt{3}} \right)
  ,\\
  \bm{b}_2 \approx \frac{2\pi}{\sqrt{3}} 
    \left( - 1 + \ce_{11} - \frac{\ce_{12}}{\sqrt{3}};\,
    \frac{1}{\sqrt{3}} + \ce_{12} - \frac{\ce_{22}}{\sqrt{3}}
  \right)
  .
\end{align}
\end{subequations}
Most importantly, the symmetry point $\bm{K}=(\tfrac{4\pi}{3\sqrt{3}},0)$
(that coincides with the Fermi point in the undoped, equilibrium situation, 
and chosen here for definiteness) moves to the new position 
\begin{equation}
  \bm{K} \approx \frac{4\pi}{3\sqrt{3}} 
    \left( 1-\frac{\varepsilon_{11}}{2}-\frac{
    \varepsilon_{22}} {2};\,-2\varepsilon_{12}\right)
  \,
  \label{eq:Shifted-K-general}
\end{equation}
for a general deformation, and in leading order in strain. For uniaxial tension
this reduces to
\begin{equation}
  \bm{K} \approx  
    \frac{4\pi}{3\sqrt{3}}\left(1-\frac{\varepsilon(1-\sigma)}{2};\,
    -\varepsilon(1+\sigma)\,\sin[2\theta]\right)
  \,.
  \label{eq:Shifted-K-tension}
\end{equation}
The factor of $2\theta$ means that the shift is the same for the \adir\ and
\zdir\ directions, in leading order. These general results will be important
for our subsequent discussion.

%
%
\section{Hopping Renormalization}
The change in bond lengths \eqref{eq:DeformedBonds} leads to
different hopping amplitudes among neighboring sites. In the Slater-Koster
scheme \cite{Slater:1954}, the new hoppings can be obtained from the
dependence of the integral $V_{pp\pi}$ on the inter-orbital distance.
Unfortunately determining such dependence with accuracy is not a trivial
matter. 
Many authors resort to Harrison's flyleaf expression which suggests that
$V_{pp\pi}(l)\propto1/l^2$ \cite{Harrison:1999}. However this is
questionable, insofar as such dependence is meaningful only in matching
the tight-binding and free electron dispersions of simple systems in 
equilibrium (beyond the equilibrium distance such dependence is unwarranted
\cite{Harrison:1999}). It is indeed known that such functional form fails away
from the equilibrium distance \cite{Grosso:1995}, and a more reasonable
assumption is an exponential decay \cite{Papaconstantopoulos:1998}.
In line with this we assume that, in graphene 
\begin{equation}
  V_{pp\pi}(l) = t_0 e^{-3.37(l/a_0-1)}
  \,,
  \label{eq:Hopping}
\end{equation}
where the rate of decay is extracted from the experimental result 
$dV_{pp\pi}/dl=-6.4\text{ eV/\AA}$ \cite{CastroNeto:2007b}. As a
consistency check, we point out that, according to Eq.~\eqref{eq:Hopping}, the
next-nearest neighbor hopping ($t'$) would have the value
$V_{pp\pi}(\sqrt{3}a_0)=0.23\text{ eV}$, which tallies with
existing estimates of $t'$ in graphene \cite{CastroNeto:2007}.

%
%
\section{Gap Threshold}

The bandstructure of Eq.~\eqref{eq:Hamiltonian} with arbitrary
hoppings $t_1,\,t_2,\,t_3$ is given by
\begin{equation}
  E(k_x,k_y) = \pm
  \left|t_2+t_3\, e^{-i\bk.\bm{a}_1}+t_1\, e^{-i\bk.\bm{a}_2}\right|
  \label{eq:Bandstructure}
  \,.
\end{equation}
Here both $t_\alpha$ and the primitive vectors $\bm{a}_\alpha$
(see \Fref{fig:Lattice} for the definition of the vectors $\bm{a}_\alpha$)
change under strain: the hoppings change as per 
Eqs.~(\ref{eq:Hopping},\ref{eq:DeformedBonds}), and the primitive vectors as
per Eq.~\eqref{eq:Gen-Vec-Deformation}.
This generalized dispersion has been previously discussed in
Refs.~\onlinecite{Wunsch:2008,Hasegawa:2006b}, under the assumption that only
the hopping elements change, without lattice deformation. It was found that the
gapless spectrum is robust, and that a gap can only appear under anisotropy
in excess of 100\% in one of the hoppings. More specifically, the spectrum 
remains gapless as long as the condition
\begin{equation}
  \left|\tfrac{|t_1|}{|t_2|}-1\right| \le \tfrac{|t_3|}{|t_2|} \le
  \left|\tfrac{|t_1|}{|t_2|}+1\right|
  \label{eq:GaplessCondition}
\end{equation}
is in effect. This condition corresponds to the shaded area in
\Fref{fig:GapThreshold}. Using the results in eqs.~(\ref{eq:StrainTensor},
\ref{eq:DeformedBonds}, \ref{eq:Hopping}) we have
mapped the evolution of the hoppings with $\ce$ and $\theta$. This allows
us to identify the range of parameters that violate
\eqref{eq:GaplessCondition}, and to obtain the threshold for gap opening.
For a given $\theta$, we follow the trajectory of the point
$(t_1/t_2,\,t_3/t_2)$ as strain grows, starting from the isotropic point at
$\ce=0$. The result is one of the arrowed curves in \Fref{fig:GapThreshold}. The
value of $\ce$ at which this curve leaves the shaded area corresponds to the gap
threshold for that particular angle. From such procedure, summarized in
\Fref{fig:GapThreshold}, we conclude that:
(i) the gap threshold is at $\ce\approx0.23$ ($\sim 20$\%);
(ii) the behavior of the system is periodic in $\theta$ with period $\pi/3$, in
accord with the symmetry of the lattice;
(iii) tension along the zig-zag direction ($\theta=0,\pi/3,\dots$) is more
effective in overcoming the gap threshold;
(iv) tension along the armchair direction never generates a gap.

The two panels of \Fref{fig:Hoppings} contain plots of the
individual $t_\alpha$ for two particular values of strain. It is clear that, for
deformations along the \zdir\ direction, the highest relative change occurs
along the zig-zag bonds ($t_{1,3}$), and conversely for deformations along the
\adir\ direction. 
This could have also been anticipated from Eqs.~\eqref{eq:DeformedBonds-ZA} and
the smallness of $\sigma$.

%
%
\begin{figure}[tb]
  \centering
  \subfigure[][]{%
    \includegraphics[width=0.33\columnwidth]{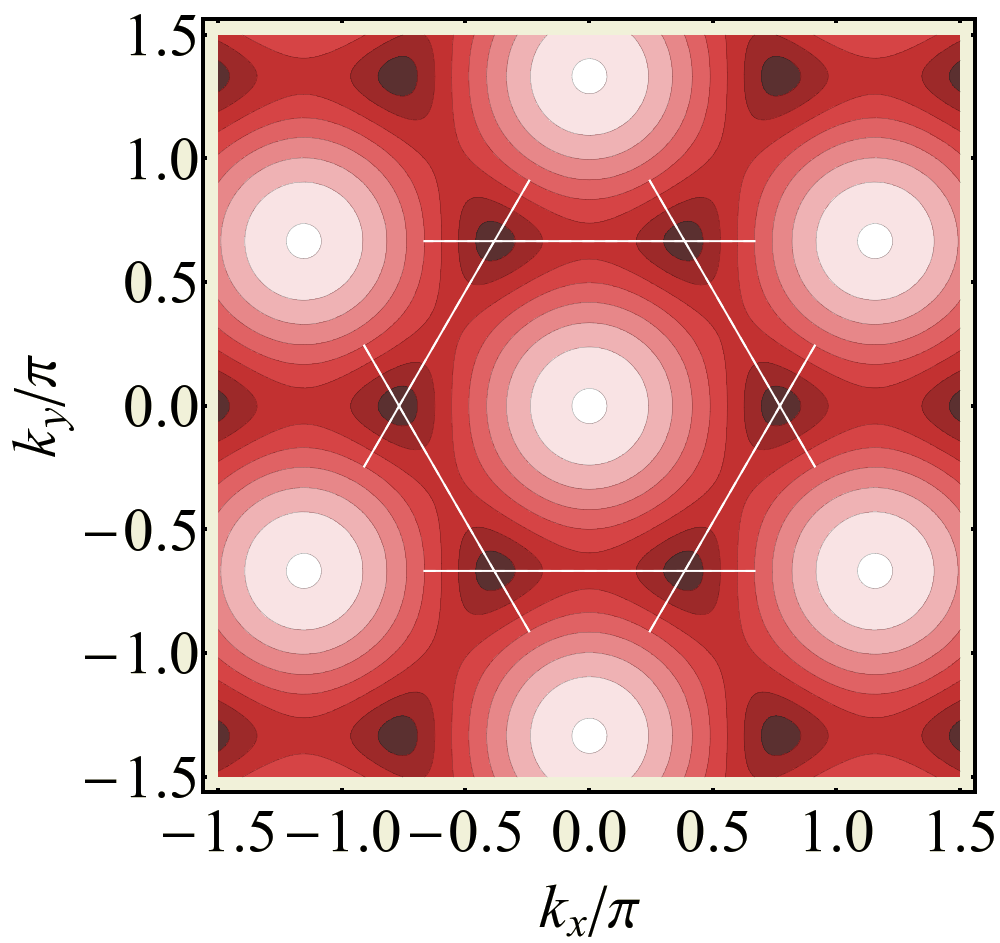}%
    \label{fig:BZ-1}%
  }%
  \subfigure[][]{%
    \includegraphics[width=0.33\columnwidth]{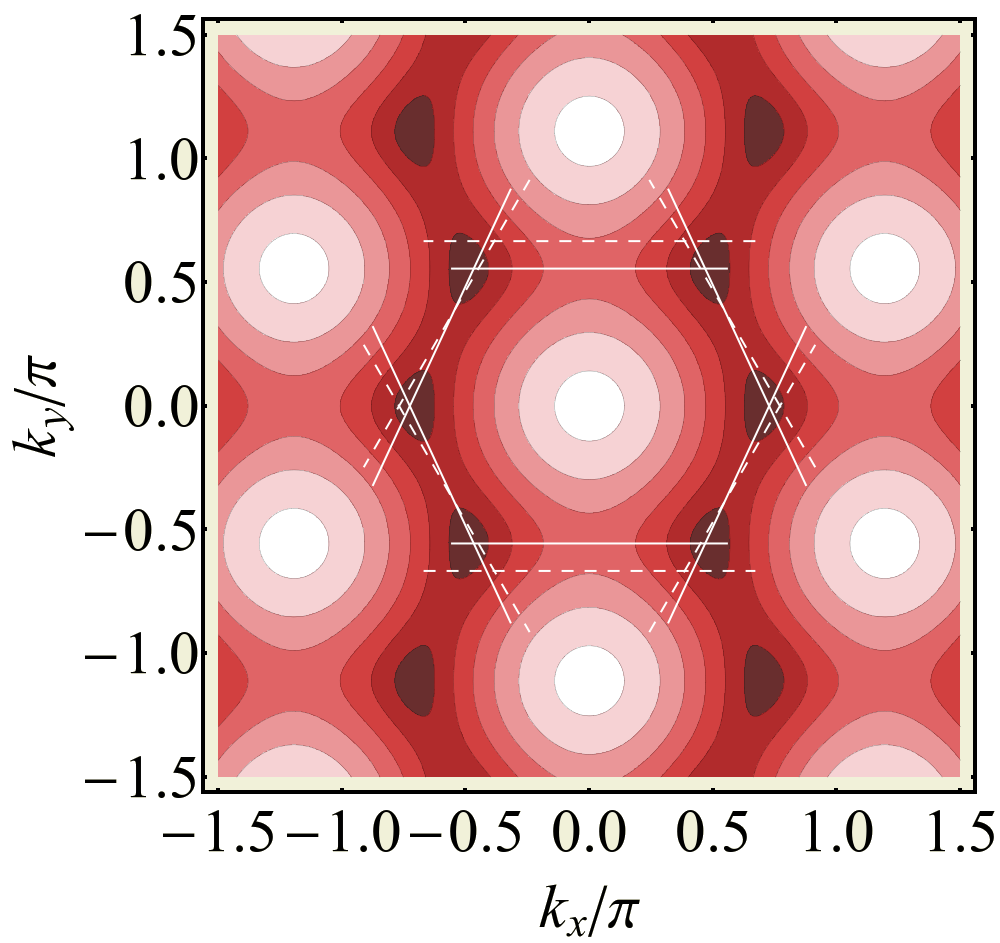}%
    \label{fig:BZ-2}%
  }%
  \subfigure[][]{%
    \includegraphics[width=0.33\columnwidth]{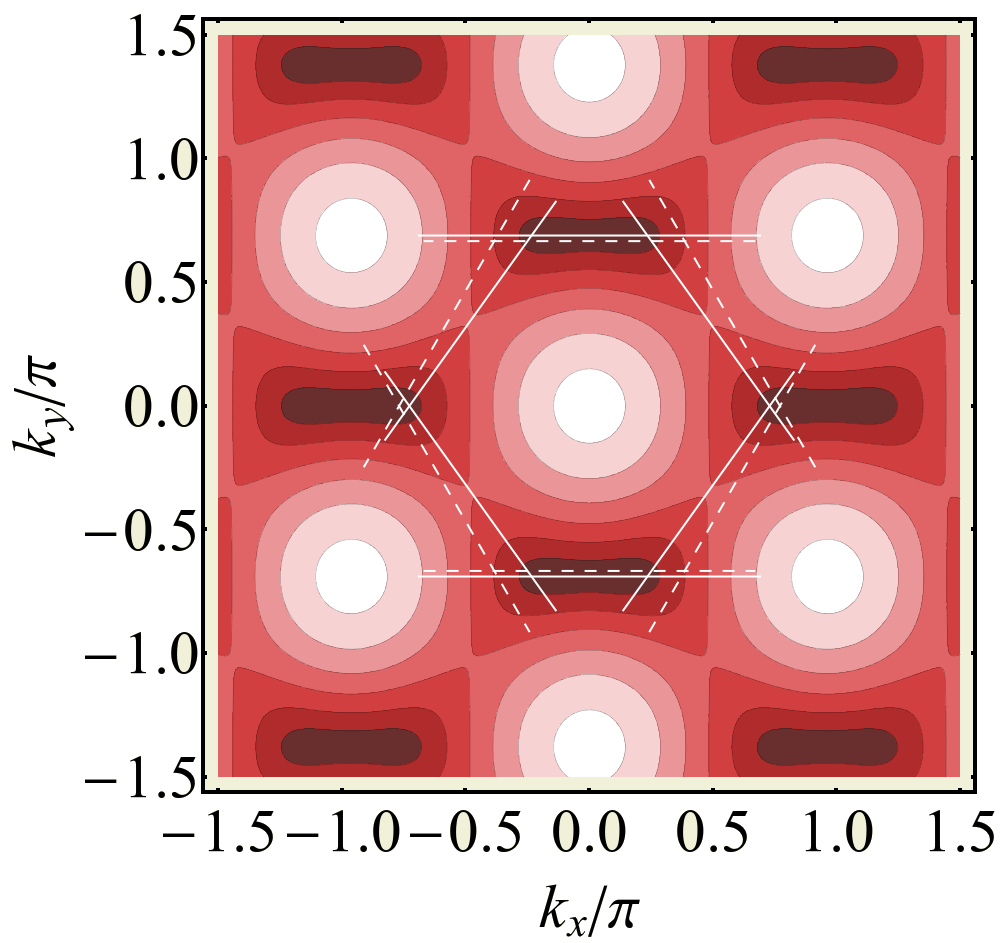}%
    \label{fig:BZ-3}%
  }\\
  \subfigure[][]{%
    \includegraphics[width=0.5\columnwidth]{%
      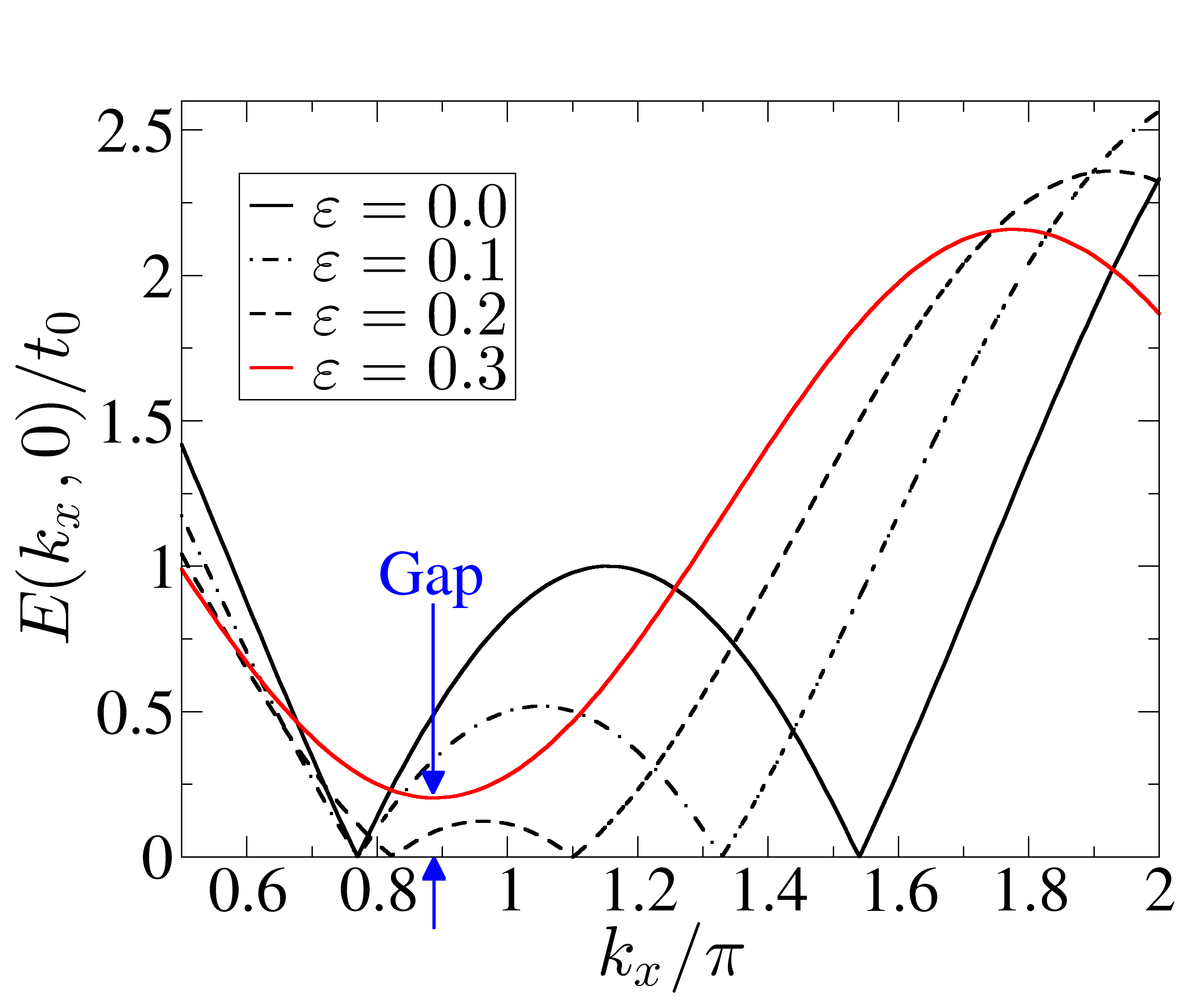}%
    \label{fig:ConeMerging}%
  }%
  \subfigure[][]{%
    \includegraphics[width=0.5\columnwidth]{%
      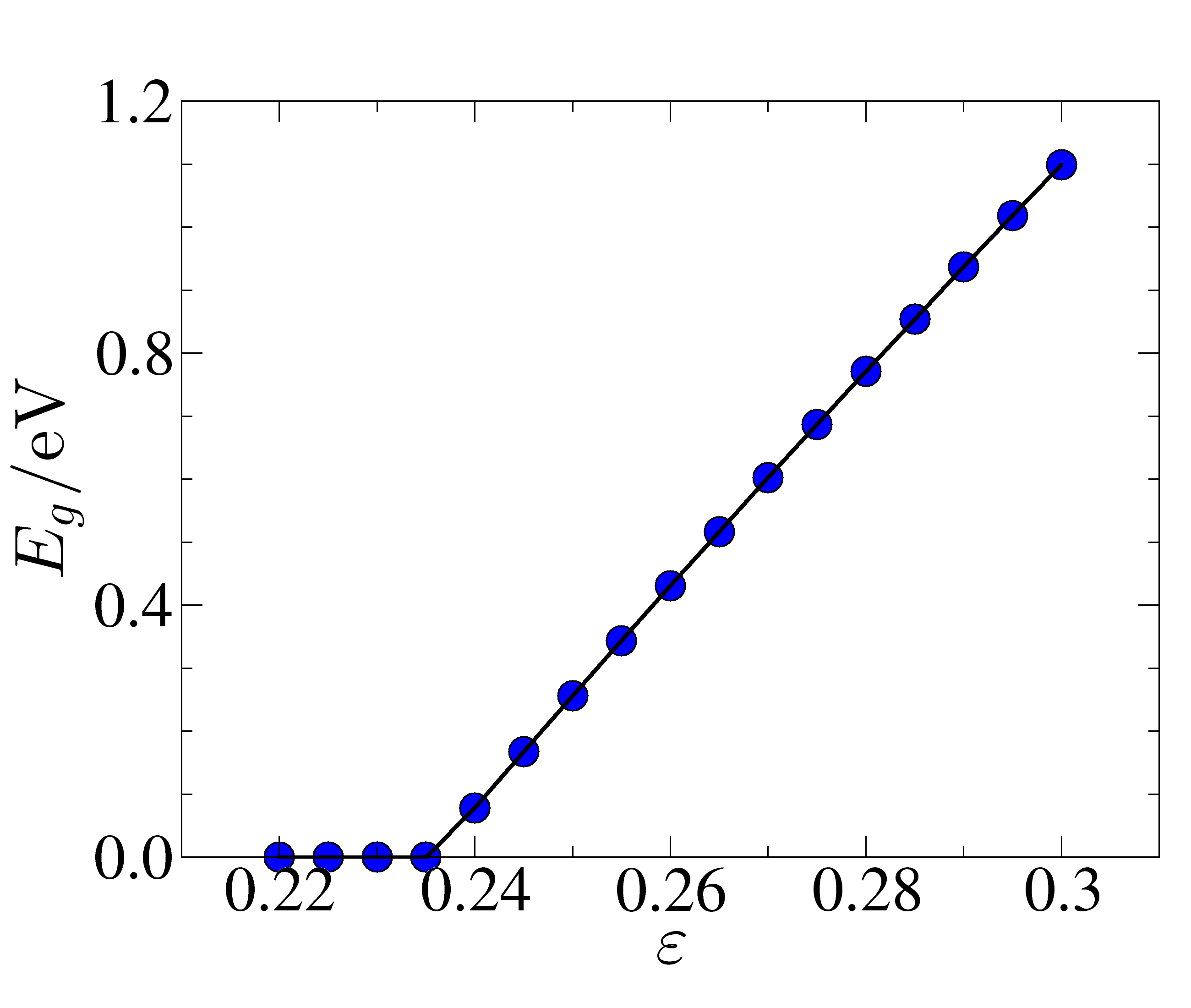}%
    \label{fig:GapVsStrain}%
  }
  \caption{
    (Color online)
    Top row shows density plots of the energy dispersion, $E(k_x,k_y)$, for 
    $\{\ce=0,\,\theta=0\}$ \subref{fig:BZ-1}, $\{\ce=0.2,\,\theta=\pi/2\}$
    \subref{fig:BZ-2}, and $\{\ce=0.2,\,\theta=0\}$ \subref{fig:BZ-3}.
    In these plots, the central white dashed lines represent the boundary of the
    first BZ of the undeformed lattice, while the solid white lines mark the 
    boundaries of the BZ for the deformed lattice.
    In \subref{fig:ConeMerging} we have a cut of \subref{fig:BZ-3}
    along $k_y=0$, showing the merging of the Dirac cones as strain increases,
    and the ultimate appearance of the gap.
    In panel \subref{fig:GapVsStrain} we compare the gap given
    by Eq.~\eqref{eq:Gap} (line) with the result obtained from direct
    minimization of the energy in the full BZ (dots).
  }
  \label{fig:BZ}
\end{figure}
%

%
%
\section{Critical Gap}
The fact that the isotropic point $(1,1)$ in \Fref{fig:GapThreshold} is
surrounded by an appreciable shaded area, means that the gapless situation is
robust, and the emergence of the gap requires a critical strain. 
The physical effect behind such critical gap lies in the fact that,
under strain, the Dirac cones drift away from the points $\bm{K},\,\bm{K}'$ in
the Brillouin zone (BZ). 
Before we proceed, it is pertinent to advance a crucial
detail: in a deformed lattice, the Dirac points (i.e. the positions in the BZ
where conduction and valence bands touch conically) and the symmetry points
$\bm{K}_i$ do not coincide. In what follows we shall distinguish them
explicitly.

To be more definite, we examine the position of the minimum energy
\cite{Endnote:6} for the bands
obtained from Eq.~\eqref{eq:Bandstructure}, which can be
done exactly if we assume that the lattice remains undeformed. Due to the
particle-hole symmetry, we minimize $E(\bk)^2$. Let us assume that $t_1=t_3\ne
t_2$, which applies for tension along either zig-zag or
armchair directions. In that case the spectrum has minima at exactly
\begin{equation}
  \bm{k}_\text{min} = 
  \Biggl(
    \pm \frac{2}{\sqrt{3}} \arccos\biggl[-\frac{t_2}{2t_1}\biggr]
    ;\, 0
  \Biggr)
  \,,
  \label{eq:k-min}
\end{equation}
and all symmetry related points. The $\pm$ sign refers to one possible choice
for the two inequivalent valleys. In addition to these local minima, that
correspond to the Dirac points, the dispersion has saddle points at 
\begin{equation}
  \bm{k} = \biggl( \frac{\pi}{\sqrt{3}} ;\, \frac{\pi}{3} \biggr)
  ,\,
  \bm{k} = \biggl(0 ;\, \frac{2\pi}{3} \biggr)
  ,\,
  \bm{k} = \biggl( \frac{\pi}{\sqrt{3}} ;\, -\frac{\pi}{3} \biggr)
  \label{eq:k-saddle}
\end{equation}
(and all symmetry related points). These are just the points
$\bm{M}_1,\,\bm{M}_2,\,\bm{M}_3$ shown in \Fref{fig:Reciprocal} and their
position is independent of $t_i$ \cite{Endnote:5}. The values of energy at these
points are 
$E(\bm{M}_1)=|t_1+t_2-t_3|$, $E(\bm{M}_2)=|-t_1+t_2+t_3|$ and
$E(\bm{M}_3)=|t_1-t_2+t_3|$.

The result in Eq.~\eqref{eq:k-min} shows that the Dirac points drift away
from the $\bm{K}$ point, and the direction of that drift is dictated by the
relative variations in $t_i$. For example, for uniaxial tension along
the \zdir\ we have $t_2>t_1=t_3$, and therefore the minimum of energy moves to
the right (left) of $\bm{K_1}$ ($\bm{K_1'}$) [cfr. \Fref{fig:Reciprocal}]. 
This means that the inequivalent Dirac points move toward each
other, and will clearly meet when $2t_1 = t_2$. They meet precisely at the
position of the saddle point $\bm{M}_2$. Throughout this process, the dispersion
remains linear along the two orthogonal directions, albeit with different Fermi
velocities. If the hoppings change
further so that $2t_1 > t_2$, the solution \eqref{eq:k-min} is no longer
valid, and the minimum lies always at $\bm{M}_2$. Since the energy at this
saddle point is given exactly by $E(\bm{M}_2)=|2t_1 - t_2|$, the system becomes
gapped, with a gap $\Delta=2|2t_1-t_2|$. Moreover, the dispersion becomes
peculiar in that it remains linear along one direction (the $y$ direction in
this example) and quadratic along the other. The topological structure
is also modified since the two inequivalent Dirac cones have merged
\cite{Montambeaux:2009}.

\begin{figure}[tb]
  \centering
  \subfigure[][]{%
    \includegraphics[width=0.45\columnwidth]{%
      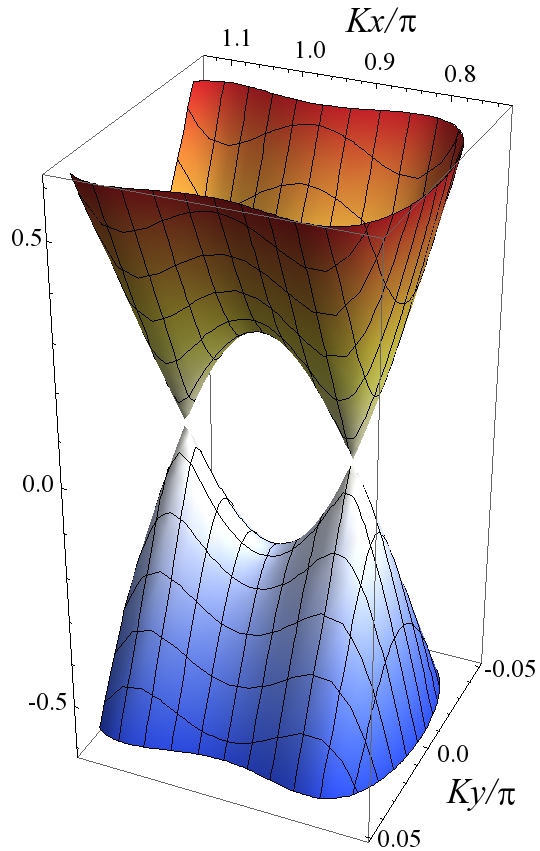}%
    \label{fig:Closeup-Precritical}%
  }%
  \subfigure[][]{%
    \includegraphics[width=0.45\columnwidth]{%
      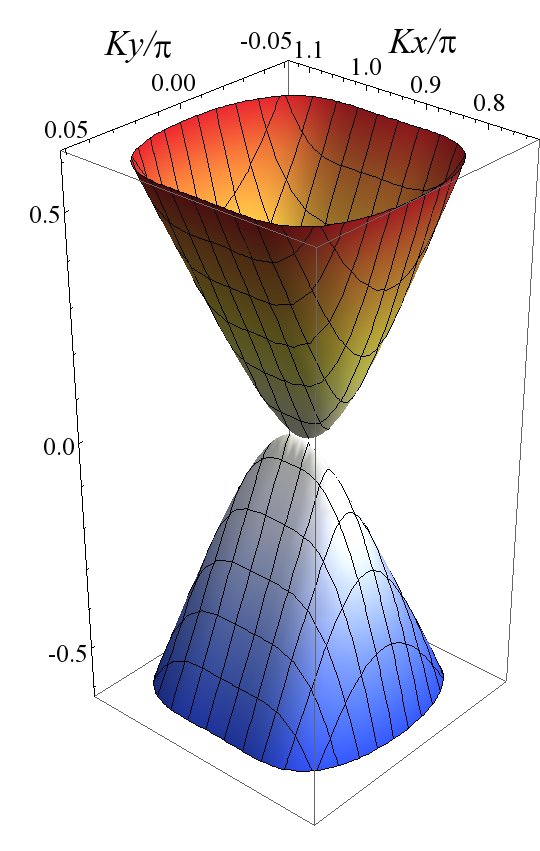}%
    \label{fig:Closeup-Critical}%
  }\\
  \subfigure[][]{%
    \includegraphics[width=0.35\columnwidth]{%
      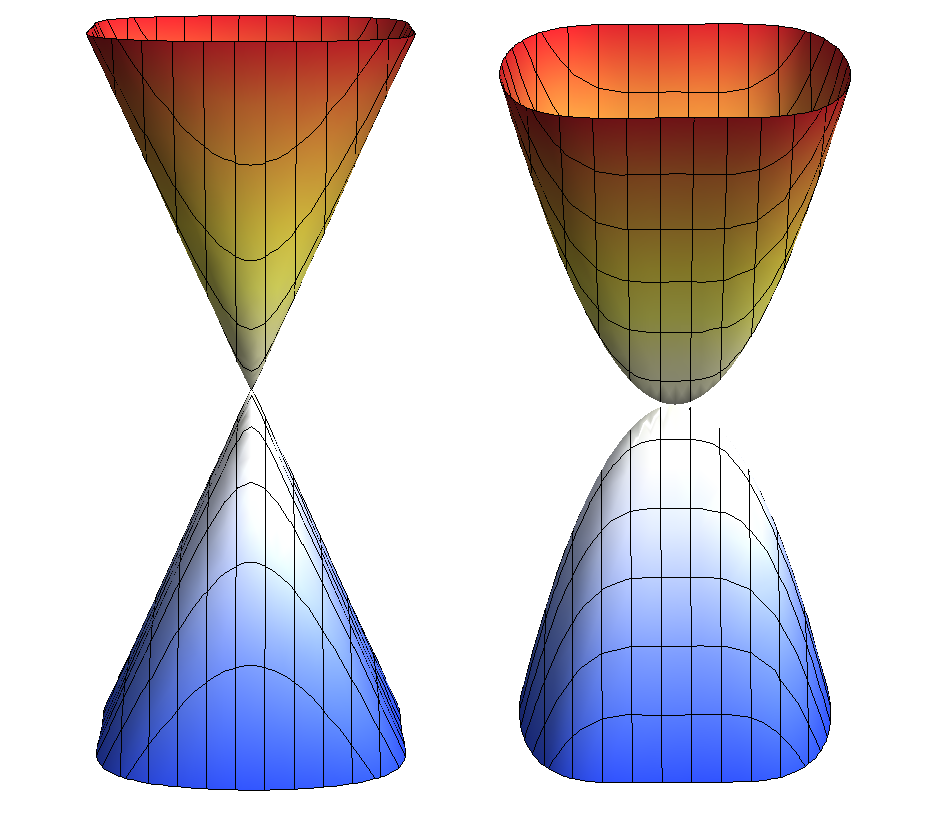}%
    \label{fig:Closeup-Critical-Angles}%
  }%
  \caption{
    (Color online)
    Closeup of the energy dispersion $E(\bk)$ in the vicinity of the Dirac 
    points. Panel \subref{fig:Closeup-Precritical} shows a pre-critical
    situation (in which the two cones persist) for a strain of $\ce=0.2$ along
    the zig-zag ($\theta=0$) direction. Notice how
    the saddle point approaches zero energy with increasing strain. In
    \subref{fig:Closeup-Critical} we show the bandstructure at precisely the
    critical strain for which the two Dirac cones meet. The dispersion is
    quadratic along $k_x$ (the direction of strain) but remains linear along
    $k_y$. This can be seen in \subref{fig:Closeup-Critical-Angles} where we
    show the critical dispersion viewed from the $k_x$ and $k_y$ axes,
    respectively.
  }
  \label{fig:Closeup}
\end{figure}

These considerations assume that the hoppings can change but the lattice
remains undeformed. Under a real
deformation both lattice and hoppings are affected. The lattice deformation
will distort the BZ but will not affect the aspects discussed above.
This can be clearly seen from inspection of the energy dispersions plotted in
\Fref{fig:BZ}(a-c). The plotted dispersions include the
deformation of the BZ and the change in the hoppings,  simultaneously. 
For strain along the \adir\ direction the nonequivalent Dirac cones move
in opposite directions and never meet [\Fref{fig:BZ-2}]. However, if
the deformation is along the \zdir\ direction, the cones always approach each
other [\Fref{fig:BZ-3}], and will eventually merge. This merging is seen in
detail in \Fref{fig:ConeMerging} where a cut along $k_y=0$ is presented to show
the emergence of the gap beyond the threshold deformation. For tension along an
arbitrary direction (except armchair) the cones always merge, the \zdir
direction being the optimal orientation, requiring less strain [cfr.
\Fref{fig:GapThreshold}].
Precisely at the critical point, the dispersion is linear along $k_x$ and
quadratic along $k_y$, as shown in \Fref{fig:Closeup}. This modification in
the dispersion along one of the Cartesian directions has peculiar implications
for the DOS and Landau level quantization \cite{Dietl:2008}.

%
%
%

The gap is a result of this Dirac cone merging process, and the origin of the
high critical strain is now clear: one needs to deform enough to bring the two
Dirac points to coincidence. This agrees with the existing understanding that
the gapless Dirac spectrum in graphene is robust (topologically protected) with
respect to small perturbations.

For strain along $\theta=0$, as discussed above, the gap is conveniently given
by 
\begin{equation}
  E_g(\ce) = 2\,\bigl|2t_1(\ce)-t_2(\ce)\bigr|\, \theta(t_2-2t_1)
  \label{eq:Gap}
  \,.
\end{equation}
An example of the strain dependence of $E_g$ can be seen in
\Fref{fig:GapVsStrain}. In it we see the agreement between the gap given by
Eq.~\eqref{eq:Gap} and the value extracted from a direct minimization of
$E(k_x,k_y)$ in the full (deformed) BZ.

From \Fref{fig:BZ-2} one can see that pulling along an armchair direction 
imparts 1D-like features to the system: the dispersion becomes highly
anisotropic. This is explained on account of the results
plotted in \Fref{fig:Hoppings} which show that stress along \adir\ tends to
weaken one bond only. In extreme cases, the weak bond can be highly suppressed
leaving only a set of 1D chains \cite{Endnote:1}. This means that strain along
certain directions can be used as a means to induce preferred anisotropy in
electric transport.
In contrast, pulling along a zig-zag direction tends to dimerize the system for
large deformations, which ultimately explains the appearance of the gap in this
case.

%
%
\begin{figure}[tb]
  \centering
  \subfigure[][]{%
    \includegraphics[width=0.5\columnwidth]{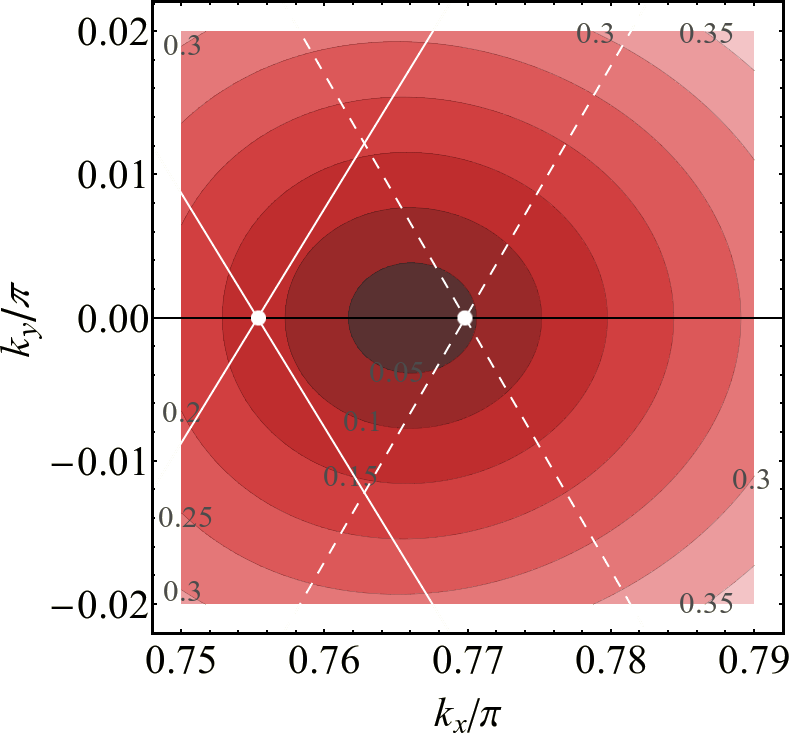}%
    \label{fig:BZ-zoom}%
  }%
  \subfigure[][]{%
    \includegraphics[width=0.5\columnwidth]{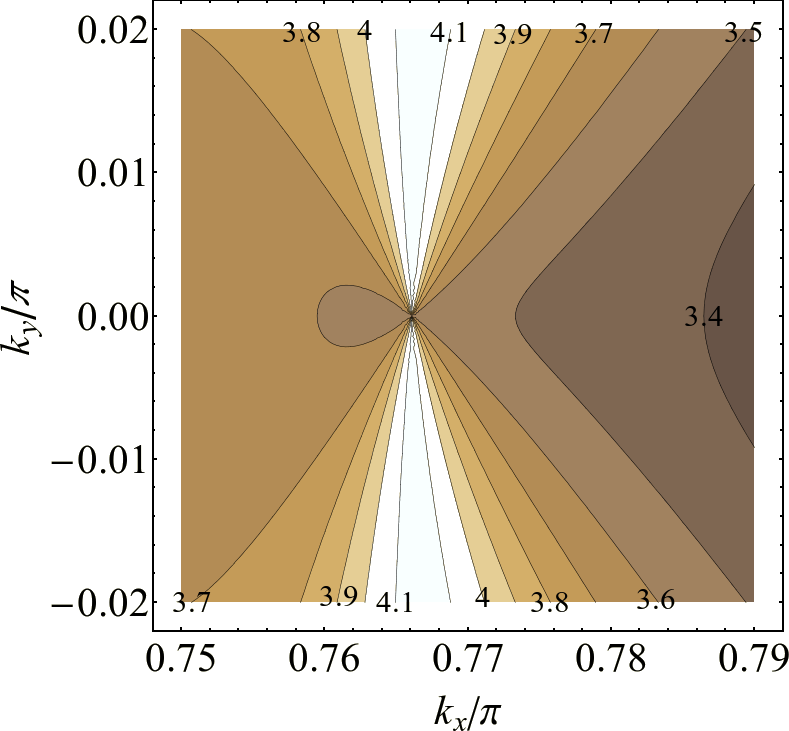}%
    \label{fig:velocity}%
  }%
  \caption{
    (Color online)
    \subref{fig:BZ-zoom}
    A close-up of the energy dispersion close to its minimum, for tension
    along $\theta=0$ and $\ce=0.05$. The solid white lines show the intersection
    of the Bragg planes that define the boundary of the first BZ in the
    deformed lattice, while the dashed white lines represent the same boundary
    es in equilibrium. It is clear that the Dirac point lies
    neither at $\bm{K}$, nor at its deformed counterpart. The energy contours
    are labeled in eV.
    \subref{fig:velocity}
    A contour plot of the absolute value of the Fermi velocity, $\hbar
    v_F=\nabla_{\bk} E(\bk)$ for the same region shown in
    \subref{fig:BZ-zoom}.
  }
  \label{fig:BZ-and-velocity}
\end{figure}
%

%
%
\section{Position of the Dirac point}
The fact that there are two concurrent effects determining the changes in the
bandstructure (viz. the lattice distortion itself, and the modification
in the nearest-neighbor hoppings) means that the position of the minimum of
energy \emph{does not coincide with the symmetry points of the deformed BZ}.
This is documented in \Fref{fig:BZ-zoom} where we provide a close-up of the
energy dispersion close to the Dirac point.
This should be clear from the foregoing discussion on the merging of the Dirac
points. In any case we want to stress this effect and illustrate it by
analyzing small perturbations with respect to the undeformed
situation. For definiteness let us focus again in the case $t_1=t_3$.
The position of the new $\bm{K}$ was given already in
eqs.~(\ref{eq:Shifted-K-general},\ref{eq:Shifted-K-tension}). The position of
the Dirac point is given by eq.~\eqref{eq:k-min} when only the hoppings change,
but not the lattice. For a small perturbation ($t_2\approx t_1$)
the result \eqref{eq:k-min} reduces to \cite{Endnote:3}
\begin{equation}
  \bm{k}_\text{min} \approx
  \pm \Biggl(
    \frac{4\pi}{3\sqrt{3}} + 2\frac{t_2-t_1}{3t_1}
    ;\, 0
  \Biggr)
  \,.
  \label{eq:k-min-2}
\end{equation}
One can calculate the correction to this result simultaneously accounting for
the lattice deformation. But the lengthy expression that results is less
important than the qualitative effect: the corrections to the expression
\eqref{eq:k-min-2} depend on the specific details of the variation of $t_i$ with
distance. Consequently, the Dirac point and the $\bm{K$} point of the deformed
lattice do not coincide in general. The equilibrium situation, in which they
coincide, is a very particular case. 

In fact, even assuming a simple lattice distortion that does not change the
hoppings will move the Dirac point away from the symmetry point of the resulting
lattice. This can be seen from a low energy expansion of
\eqref{eq:Bandstructure} putting $t_i=t$. The position of the Dirac point
$\bm{K}_1$ in such case moves to
\begin{equation}
  \bm{k}_\text{min} \approx
  \Biggl(
    \frac{4\pi}{3\sqrt{3}} (1-\ce_{11})
    ;\, -\frac{4\pi}{3\sqrt{3}} \ce_{12}
  \Biggr)
  \,,
\end{equation}
which is clearly different from \eqref{eq:Shifted-K-general}. This fact is of
critical relevance when interpreting results of similar calculations obtained
\emph{ab-initio}, as will be discussed below.

The fact that the Dirac point drifts from the corner of the BZ means that there
are no longer 3 equivalent pairs of points in the first BZ for the
neutral system, but only one pair of non-equivalent points in general (in other
words, in the undoped and undeformed lattice the Fermi surface is distributed
among the 6 degenerate K points in the boundary of the BZ,
whereas for a general deformation we have only two within the ﬁrst BZ).
In the situation shown in \Fref{fig:BZ-zoom}, for example, the Dirac point shown
in the figure lies outside the first Brillouin zone. The one inside the first BZ
is actually the (equivalent) Dirac point that moved away from $\bm{K}_2$ (or
$\bm{K}_3$), in the notation of \Fref{fig:Reciprocal}.

%
%
\section{Discussion}

We have seen that, within the tight-binding Hamiltonian
written in Eq.~\eqref{eq:Hamiltonian}, uniform tension can induce
a bulk spectral gap in graphene. However, at least within a non-interacting
tight-binding approach, the gap threshold is very difficult
to overcome, if at all possible. Since a tensional strain in excess of 20\% is
required to observe such feature, several comments are in order.

\subsection{On the approximations employed}

We start by noticing that in our calculation we kept only the lowest order
terms in $\ce$. In addition, although strain magnitudes of $\sim 20$\% are not
unreasonable, graphene is expected to be in the non-linear elastic
regime at those deformations \cite{Liu:2007}. Therefore, non-linear
corrections can be relevant at the quantitative level in the vicinity of the
threshold.

Notwithstanding, our main result is robust: no gap can be opened under
planar tension situations, except in highly strained situations. This
conclusion does not depend on having taken a linear approximation insofar as it
should be valid up to deformations in the range of 5-10\%.

With respect to our tight-binding parametrization including only nearest
neighbor hopping, we should mention that Kishigi \etal
have shown that inclusion of next-nearest neighbor terms ($t'$) can, alone,
generate a gap \cite{Kishigi:2008}. But this requires a very specific
deformation of the lattice, unlikely to occur under simple tension. 
The presence of $t'$ can also lead to other effects, like tilted Dirac 
cones as discussed in Ref.~\onlinecite{Goerbig:2008}. 

It is expected that the planar arrangement of carbon atoms in freely hanging
graphene should become unstable with respect to a buckled or rippled
configuration, or even experience mechanical failure for moderate to high
tension. The presence of a substrate should provide more stability for the
planar distribution of the carbon atoms. In fact, a recent experiment
published during revision of this manuscript \cite{Kim:2009} shows that
reversible strain of the order of 18\% can be induced in graphene deposited on
flexible plastic substrates. 

\subsection{On related \emph{ab-initio} calculations}

Secondly, some \emph{ab-initio} calculations seem to show that a gap is present
in graphene for arbitrarily small tensions \cite{Ni:2008b,Gui:2008}. But these
reports have some conflicting details. For example there is an order of
magnitude discrepancy between the gap predicted in these two references for 1\%
strain. In addition, Ref.~\onlinecite{Gui:2008} claims their \emph{ab-initio}
result agrees with the bandstructure \eqref{eq:Bandstructure} after a suitable
choice of hoppings. As we showed above this cannot be the case, since there is
always a (large) threshold for the appearance of the gap. Consequently, further
clarification regarding \emph{ab-initio} under strain is desired.

One issue that requires special attention in interpreting density functional
theory (DFT) 
calculations of graphene's bandstructure under strain is the shift of the Dirac
point. As we stressed earlier, when graphene is strained the Dirac point
(position of the energy minimum) does not lie at any symmetry point of the
lattice. This is paramount because DFT calculations of the bandstructure rely
on a preexisting mesh in reciprocal space, at whose points the bandstructure is
sampled. These meshes normally include points along high symmetry lines of the
BZ. But in the current problem, the use of such a traditional mesh will not be
particularly useful to distinguish between a gapped and gappless situation.
Since the Fermi surface of undoped graphene is a single point, unless the
sampling mesh includes that precise point, one will always obtain a gap in the
resulting bandstructure. 

Most recently, we became aware of two new developments from the
\emph{ab-initio} front, that shed the light needed to interpret the earlier
calculations mentioned above. One of them \cite{Ni:2009}
is a revision of the DFT calculations presented in Ref.~\onlinecite{Ni:2008b}.
In this latter work it is shown that, upon careful analysis, the DFT calculation
shows no gap under uniaxial deformations up to $\sim 20\%$. In fact the
authors mention explicitly that the shift of the
Dirac point from the high symmetry point mislead the authors in their initial
interpretation of the bandstructure. In another preprint \cite{Farjam:2009},
independent authors show that their DFT calculations reveal,
again, no gap for deformations of the order of 10\% in either the \zdir\ or
\adir\ directions. These subsequent developments confirm our prediction that
only excessive planar deformations are able to engender a bulk spectral gap in
graphene.

\subsection{Anisotropic Transport}

Several effects of tensional strain are clear from our results. Tension leads
to one-dimensionalization of transport in graphene by weakening preferential
bonds: transport should certainly be anisotropic, even for small tensions. One
example of that is seen in \Fref{fig:BZ-zoom} where a strain of 5\% visibly
deforms the Fermi surface. \Fref{fig:velocity}, where the Fermi velocity is
plotted for the same region in the BZ, further shows that, for the chosen
tension direction, the Fermi surface is not quite elliptical but slightly oval,
in a reminiscence of the trigonal warping effects.

\begin{figure}[tb]
  \centering
 
\includegraphics[width=0.8\columnwidth]{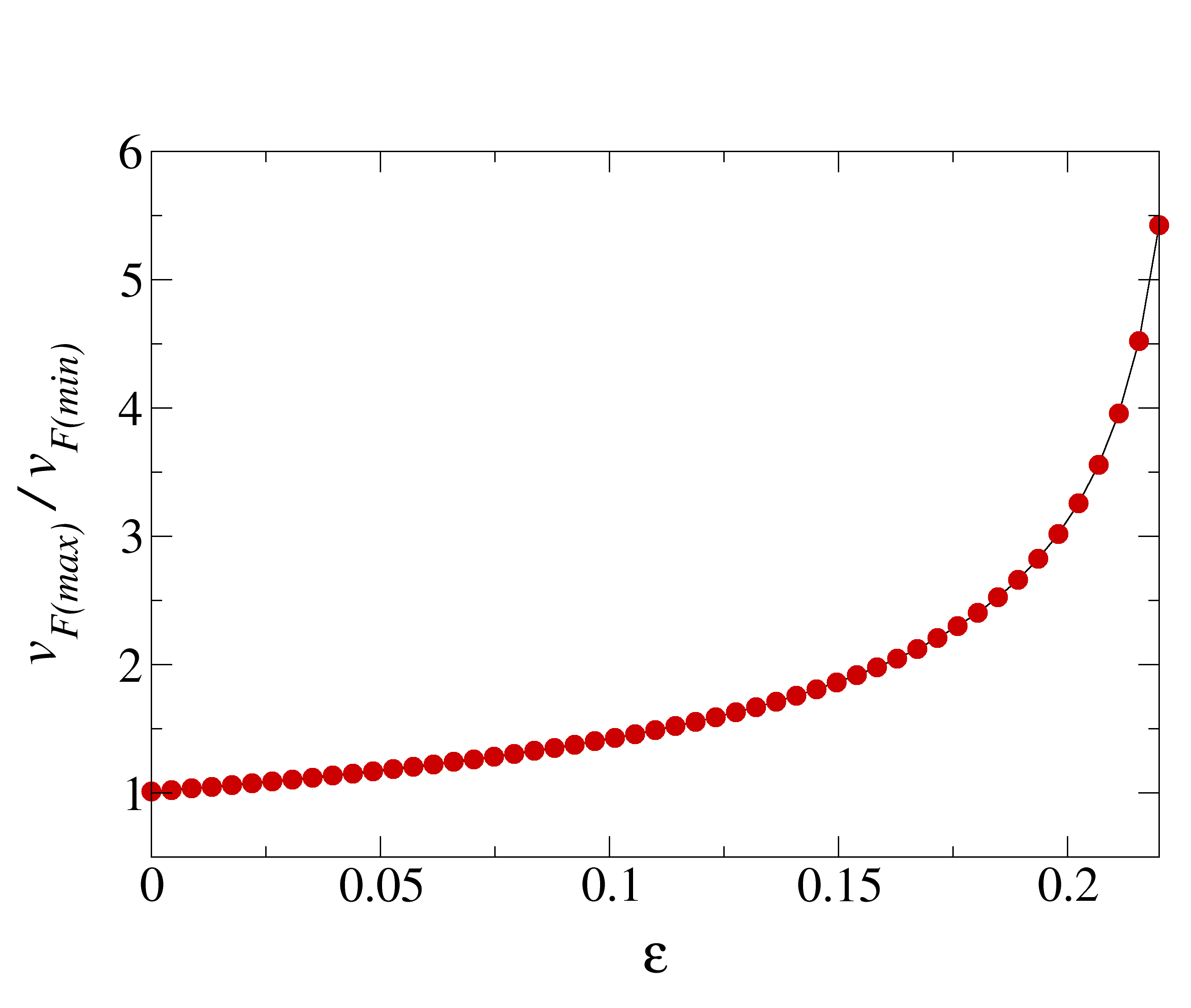}%
  \caption{
    (Color online)
    Anisotropy in the Fermi velocity, as a function of strain for a Fermi
    energy of 50meV. The vertical axis shows the ratio of the maximum to the
    minimum values of $v_F$ in the entire Fermi surface.
  }
  \label{fig:Vf-anisotropy}
\end{figure}

The anisotropy in the Fermi velocity can become quite large, as shown in
\Fref{fig:Vf-anisotropy}. There we plot the ratio
$v_{F\text{(max)}}/v_{F\text{(min)}}$ for
$E_F=50$~meV. 
As can be seen in the cuts of \Fref{fig:ConeMerging}, near the critical strain
the Fermi level may touch the van Hove singularity (midpoint between the two
cones) where $v_F = 0$ along one direction, which leads to formal divergence of
the ratio between the maximum and minimum Fermi velocities.

The Fermi surface anisotropy has been captured in transport
experiments that reveal a considerable anisotropy in the resistivities parallel
($R_{xx}$) and perpendicular ($R_{yy}$) to the tension direction
\cite{Kim:2009}. The authors of
Reference \onlinecite{Kim:2009} report in their Fig.~4c resistivity
anisotropies of up to one order of magnitude at $\sim 19\%$ strain. We can make
a simple estimate of the anisotropies expected in the light of our results for
the deformed bandstructure. For that we follow a Boltzmann approach to the
relaxation time and DC conductivity. As is usually the case, the relevant
electron states are the ones lying in a narrow vicinity of the Fermi surface.
On account of the Fermi surface anisotropy, we can take the Fermi velocities in
the direction of the electric field only. For example, in the situation shown in
\Fref{fig:BZ-zoom} for uniaxial strain along $Ox$, the relevant $v_F$'s
determining $R_{xx}$ and $R_{yy}$ will be the ones along the major and minor
axes, respectively. 

For the purposes of our estimate, we take the Boltzmann longitudinal
conductivity for scattering out of unscreened charged impurities of valence $Z$
and concentration $n_i$ \cite{Peres:2007}, given by
\begin{equation}
  \sigma = \frac{2e^2}{h}\frac{\pi\hbar^2v_F^2n}{Z^2e^2n_i}
  \,,
  \label{eq:Boltzmann}
\end{equation}
and follow the reasoning above: replacing $v_F\to\bm{v}_F.\bm{u}_E$,
where $\bm{u}_E$ is the direction of the electric field in each case (parallel
or perpendicular to the tension direction). We immediately see that this
estimate leads to a resistance anisotropy of \cite{Endnote:7}
\begin{equation}
  \frac{R_{xx}}{R_{yy}} \approx 
    \biggl(\frac{v_{F\text{(max)}}}{v_{F\text{(min)}}} \biggr)^2
  .
  \label{eq:Resistance}
\end{equation}
For the maximum strain used in the experiment ($\sim19\%$) our plot in
\Fref{fig:velocity} yields a ratio of $2.8$, which, per \eqref{eq:Resistance},
corresponds to a resistance anisotropy of roughly 8 fold, consistent with the
measured anisotropy.

Our results apply to exfoliated and epitaxial graphene alike. As matters
currently stand, it is perhaps more relevant in the context of the latter, since
graphene grown epitaxialy on SiC is almost always under strain
\cite{Endnote:4}. The strained configurations are imposed by the lattice
mismatch with the substrate, and can be controlled by changing the growth and
annealing conditions \cite{Ferralis:2008}. For these systems, the relaxed
starting configuration is already deformed.

Lastly, it is important to point out that, even though a spectral gap
seems to require extreme strain, one can generate a \emph{transport} gap
by means of local, small, deformations. It has been shown in
Refs.~\onlinecite{Pereira:2008,Fogler:2008} that tunneling across a
locally strained region is highly suppressed, and leads to a transport gap
(i.e.,
a suppression of  electrical conductivity) at small densities, even in the
absence of a bulk spectral gap. 

%
%
%

%
%
\section{Conclusions}
Within a non-interacting, nearest-neighbor tight-binding approach, we have shown
that opening a spectral gap in strained graphene requires deformations of the
order of 20\%. This result is confirmed by the most recent \emph{ab-initio}
calculations. Such an extreme strain is required on account of the stability of
the Dirac points in graphene, that renders the spectrum gappless unless the two
inequivalent Dirac points merge. The merging requires substantial
anisotropy in the hopping integrals, only achieved under high strain. General
features of strained graphene are an anisotropic Fermi surface, anisotropic
Fermi velocities, and the drift of the Dirac points away from the high symmetry
points of the lattice.

Uniform planar stain appears to be an unlikely candidate to induce a bulk gap in
graphene. Nevertheless, strain (local or uniform) can be an effective
means of tuning the electronic structure and transport characteristics
of graphene devices. Even if the bulk gap turns out to be challenging in
practice, local strain could be used as a way to mechanically pinch off current
flow.

%
%
\acknowledgments
We thank enlightening discussions with Y. P. Feng, A. K. Geim, A. Heinz, Y.
Lu, Z. Ni, Z.~X. Shen, and T. Yu. 
AHCN acknowledges the partial support of the U.S. Department of Energy
under the grant DE-FG02-08ER46512.
VMP is supported by FCT via SFRH/BPD/27182/2006 and 
PTDC/FIS/64404/2006. 


\bibliographystyle{apsrev}
\bibliography{graphene_strain}

\begin{thebibliography}{41}
\bibitem[{End({\natexlab{a}})}]{Endnote:2}
\bibinfo{note}{In this case the relation between strain and stress is given by
  the material parameters of the substrate, rather than the intrinsic elastic
  parameters for graphene.}
\bibitem[{End({\natexlab{b}})}]{Endnote:6}
\bibinfo{note}{By ``position of the minimum energy'' we mean the position of
  the Dirac points (or valleys). They are local extrema, with the abolute
  minimum of energy lying at the $\Gamma$ point in the BZ.}
\bibitem[{End({\natexlab{c}})}]{Endnote:5}
\bibinfo{note}{The position of the saddle points does not depend explicitly
  upon the hoppings $t_i$, and their position is constant only if the lattice
  remains undeformed.}
\bibitem[{End({\natexlab{d}})}]{Endnote:1}
\bibinfo{note}{This is another way to see why strain along A can never lead to
  a gap.}
\bibitem[{End({\natexlab{g}})}]{Endnote:4}
\bibinfo{note}{Although in this case the most common situation is biaxial
  strain}.
\end{thebibliography}

\end{document}